\documentclass[aps,pra,twocolumn,superscriptaddress]{revtex4-2}
\usepackage{graphicx}
\graphicspath{{../},{figures/}}
\usepackage{amssymb}
\usepackage{comment}
\usepackage{amsmath}
\usepackage[utf8x]{inputenc}
\usepackage[T1]{fontenc}
\usepackage{dsfont}
\usepackage{url}
\usepackage[colorlinks]{hyperref}
\hypersetup{%
    plainpages=true,
    breaklinks=true,
    hypertexnames=false,
    pageanchor=true,
    colorlinks=true,
    linkcolor={blue},
    citecolor={red},
    urlcolor={blue},
    anchorcolor={black}
    }

\usepackage[position=top,caption=false]{subfig}

\newcommand{\ket}[1]{| #1 \rangle}
\newcommand{\bra}[1]{\langle #1 |}

\def\tr{{\rm Tr}}

\def\RE{{\rm Re}}
\def\IM{{\rm Im}}

\def\CP{{\rm CP}}
\def\NP{{\rm NP}}
\def\SP{{\rm SP}}
\def\BP{{\rm BP}}

\def\GCP{{\rm CP}_{qr}}
\def\GNP{{\rm NP}_{qr}}
\def\GSP{{\rm SP}_{qr}}
\def\GBP{{\rm BP}_{qr}}

\def\<{\langle}
\def\>{\rangle}

\begin{document}



\title{Quantifying nonclassicality of vacuum-one-photon
superpositions via potentials for\\ Bell nonlocality, quantum
steering, and entanglement}

\author{Adam Miranowicz} 
 \affiliation{Institute of Spintronics and Quantum Information, Faculty of Physics, Adam Mickiewicz University, 61-614 Pozna\'n, Poland}
 \affiliation{Theoretical Quantum Physics Laboratory, Cluster for Pioneering Research, RIKEN, Wakoshi, Saitama 351-0198, Japan}

\author{Josef Kadlec} 
 \affiliation{Joint Laboratory of Optics of Palack\'y University and Institute of Physics of Czech Academy of Sciences, 17. listopadu 12, 779 00 Olomouc, Czech Republic}

\author{Karol Bartkiewicz} 
 \affiliation{Institute of Spintronics and Quantum Information, Faculty of Physics, Adam Mickiewicz University, 61-614 Pozna\'n, Poland}

\author{Anton\'in \v{C}ernoch} 
 \affiliation{Institute of Physics of the Czech Academy of Sciences, Joint Laboratory of Optics of PU and IP AS CR, 17. listopadu 50A, 772 07 Olomouc, Czech Republic}

\author{Yueh-Nan Chen}
  \affiliation{Department of Physics, National Cheng Kung University, Tainan 70101, Taiwan}
  \affiliation{Center for Quantum Frontiers of Research \& Technology, NCKU, Tainan 70101, Taiwan}

\author{Karel Lemr} 
 \affiliation{Joint Laboratory of Optics of Palack\'y University and Institute of Physics of Czech Academy of Sciences, 17. listopadu 12, 779 00 Olomouc, Czech Republic}

\author{Franco Nori} 
    \affiliation{Theoretical Quantum Physics Laboratory, Cluster for Pioneering Research, RIKEN, Wakoshi, Saitama 351-0198, Japan}
    \affiliation{Center for Quantum Computing, RIKEN, Wakoshi, Saitama 351-0198, Japan}
    \affiliation{Department of Physics, University of Michigan, Ann Arbor, Michigan 48109-1040, USA}

\begin{abstract}
Entanglement potentials are popular measures of the
nonclassicality of single-mode optical fields. These potentials
are defined by the amount of entanglement (measured by, e.g., the
negativity or concurrence) of the two-mode field generated by
mixing a given single-mode field with the vacuum on a balanced
beam splitter. We generalize this concept to define the potentials
for Bell nonlocality and quantum steering in specific measurement
scenarios, in order to quantify single-mode nonclassicality in a
more refined way. Thus, we can study the hierarchy of three types
of potentials in close analogy to the well-known hierarchy of the
corresponding two-mode quantum correlations. For clarity of our
presentation, we focus on the analysis of the nonclassicality
potentials for arbitrary vacuum-one-photon superpositions (VOPSs),
corresponding to a photon-number qubit. We discuss experimentally
feasible implementations for the generation of single-mode VOPS
states, their mixing with the vacuum on a balanced beam splitter,
and their two-mode Wigner-function reconstruction using homodyne
tomography to determine the potentials. We analyze the effects of
imperfections, including phase damping and unbalanced beam
splitting on the quality of the reconstructed two-mode states and
nonclassicality potentials. Although we focus on the analysis of
VOPS states, single-mode potentials can also be applied to study
the nonclassicality of qudits or continuous-variable systems.
\end{abstract}
\date{\today}

\maketitle

\section{Introduction}

Nonclassical optical states (including entangled, squeezed, or
photon antibunched) are the main resources for quantum
technologies and quantum information processing with photons.
Thus, testing and quantifying the nonclassicality (NC) of optical
states has been attracting attention in quantum physics since the
pioneering work of Kennard~\cite{Kennard1927} on squeezed states
published almost a century ago. It is worth noting that the first
truly convincing experimental demonstration of the nonclassical
character of photons was based on measuring photon
antibunching~\cite{Kimble1977}. Recent experimental optical
demonstrations of quantum advantage include enhanced
gravitational-wave detection with squeezed
states~\cite{Abadie2011,Aasi2013,Grote2013}, boson sampling based
on entangled and squeezed states~\cite{Zhong2020, Zhong2021,
Madsen2022}, and entanglement-based quantum
cryptography~\cite{Yin2020}.

In quantum optics, the state of an optical field is classified as
nonclassical (or quantum) if its Glauber-Sudarshan $P$
function~\cite{GlauberBook,Sudarshan1963} is not positive
semidefinite, so it is not a classical probability
density~\cite{ScullyBook, AgarwalBook, VogelBook}. This means that
only coherent states and their statistical mixtures are considered
classical. Extensive attention has been devoted to various forms
of nonclassical correlations, with particular emphasis on their
three distinct types: quantum entanglement (quantum
inseparability)~\cite{HorodeckiReview}, Einstein-Podolsky-Rosen
(EPR) steering (commonly referred to as quantum
steering)~\cite{CavalcantiReview, UolaReview}, and Bell
nonlocality, which manifests through violations of Bell
inequalities~\cite{BrunnerReview}. In this paper, we quantify the
NC of single-qubit optical states, which are arbitrary vacuum and
one-photon superpositions (VOPS), via measures of two-mode quantum
correlations. Testing nonlocal quantum correlations of
single-photon states, or more specifically the states, generated
by mixing a VOPS with the vacuum on a balanced beam splitter (BS),
has been attracting considerable interest both theoretical (see,
e.g.,~\cite{Tan1991, Hardy1994, Jones2011, Kogias2015}) and
experimental (see, e.g.,~\cite{Lombardi2002, Lvovsky2002,
Babichev2004, Fuwa2015}).

Experimental tests whether a given optical state is nonclassical
are usually based on measuring NC witnesses corresponding to
demonstrating violations of various classical
inequalities~\cite{VogelBook, DodonovBook, PerinaBook, Adam2010,
Bartkowiak2011}. Typical NC witnesses are not universal, which
means that they are sufficient but not necessary criteria of NC.
Universal witnesses of NC correspond to those criteria which are
both sufficient and necessary of NC. Experimental implementations
of such universal witnesses usually require applying a complete
quantum state tomography (QST). They can be used not only as NC
tests but also NC measures (if  they satisfy some additional
properties). The most popular NC measures include: nonclassical
distance~\cite{Hillery1987}, nonclassical depth~\cite{Lee1991,
Lutkenhaus1995}, and entanglement potentials~\cite{Asboth2005}.

Nonclassical depth is defined as the minimal amount of Gaussian
noise which transforms a nonpositive semidefinite $P$ function
into a positive one. It is equal to 1 for all non-Gaussian states,
and, thus, it is not a useful measure to quantify the amount of
the NC of VOPS states~\cite{Adam2015a}, although it was shown to
be useful for quantifying the NC of Gaussian states, e.g., twin
beams~\cite{Arkhipov2015}. Moreover, nonclassical distance is
defined as the distance (according to a chosen distance measure
including those of Bures or Kullback-Leibler) of a given
nonclassical state to its closest classical state (CCS). Finding a
CCS is usually very hard, even numerically. Of course, if one
limits the set of classical states, then the nonclassical distance
can be calculated effectively; e.g., it can be calculated for VOPS
states if the vacuum is chosen as the CCS, which is reasonable
because this is the only classical VOPS state~\cite{Adam2015a}. Of
course, the CCS for a given VOPS state might belong to a wider
class of classical states. So, in general, finding a CCS could be
difficult even for such simple VOPS states. Thus, we consider here
only entanglement potentials and related NC quantifiers which do
not suffer from the above-mentioned problems of nonclassical depth
and distance.

Non-universal NC witnesses, which are also often used for
quantifying NC, to mention only a few include: (i) quadrature
squeezing variances; (ii) second-order correlation functions to
quantify photon antibunching and the sub-Poissonian photon
statistics; (iii) the nonclassical volume corresponding to the
volume of the negative part of a Wigner
function~\cite{Kenfack2004}; (iv) the Wigner distinguishability
quantifier~\cite{Mari2011}, which is defined in terms of the
distinguishability of a given state from one with a positive
Wigner function; and (v) quantifiers of two- and multi-mode
quantum correlations, which are the main topic of this paper, can
also be used for estimating the degree of
NC~\cite{Asboth2005,Vogel2014,Adam2015a,Adam2015b,Killoran2016}.
We also mention operational approaches to quantify the NC of
states (see, e.g.,~\cite{Gehrke2012,Meznaric2013,Nakano2013})
including the effect of a measurement setup. For example, the
negativity of quantumness is defined as the minimum entanglement
(quantified by the negativity) that is created between a given
system and a measurement apparatus assuming local measurements
performed on subsystems~\cite{Nakano2013}.

The well-known hierarchy of standard measures of entanglement, EPR
steering (in different measurement scenarios), and Bell
nonlocality has recently been demonstrated for experimental
polarization-qubit states,  which were measured by applying
complete~\cite{Jirakova2021, Ku2022} or incomplete~\cite{Abo2023}
QST. A closely related hierarchy of temporal quantum correlations,
including temporal inseparability, temporal steering, and
macrorealism, has also been studied~\cite{Ku2018}. Moreover,
considerable research has been devoted to the hierarchies of
measures or witnesses of NC, which are limited to specific types
of quantum correlations. These include hierarchies of entanglement
witnesses~\cite{Shchukin2005,Adam2006}, steering
witnesses~\cite{Kogias2015}, Bell
inequalities~\cite{Navascus2007,BrunnerReview}; as well as
spatial~\cite{Richter2002} and spatiotemporal~\cite{Vogel2008,
Adam2010} NC witnesses.

In this paper we study theoretically \emph{the potentials for
two-qubit correlations to quantify the NC of single-qubit states
defined as (coherent or incoherent) VOPSs.} We focus on analyzing
quantifiers of single-qubit NC based on the above-mentioned three
types of quantum correlations, i.e., entanglement, steering, and
Bell nonlocality. More specifically, inspired by the concept of
entanglement potentials introduced in Ref.~\cite{Asboth2005} for
quantifying the NC of single-mode optical states, \emph{we
introduce the potentials for EPR steering and Bell nonlocality.
These potentials for two-mode quantum correlations can serve as
the quantifiers of single-mode NC correlations.} In particular,
they can also enable us to determine the hierarchy of single-qubit
nonclassical correlations via the corresponding hierarchy of
two-qubit nonclassical correlations.

Compared to our former related works on quantifying NC of
single-qubit optical states~\cite{Adam2015a,Adam2015b}, here we
introduce \emph{novel types of potentials for two-qubit
correlations to study the hierarchy of single-qubit
nonclassicality,} analogously to the hierarchy of two-qubit
correlations, which we studied experimentally in
Refs.~\cite{Jirakova2021, Ku2022, Abo2023} using
polarization-based tomographic methods (see
Refs.~\cite{Miranowicz2014tomo, Bartkiewicz2016tomo} for
comparative analyses).

\begin{figure}[hbt!]
{\includegraphics[width=0.8\columnwidth]{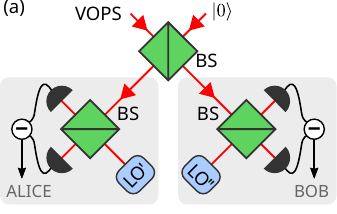}}\hspace*{0pt}\\ %
{\includegraphics[width=0.8\columnwidth]{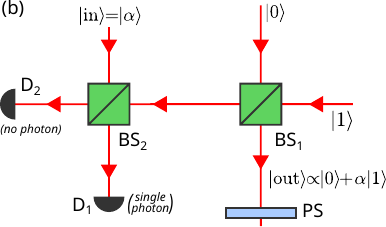}}\hspace*{0pt} %
\caption{(a) Scheme for converting the nonclassicality of a
vacuum-one-photon superposition (VOPS) state $\sigma(p,x)$ into a
two-mode state $\rho(p,x)$ exhibiting entanglement, and in some
cases EPR steering and Bell nonlocality; $\rho(p,x)$ can be
reconstructed by homodyne state tomography, where LO' and LO''
stand for local oscillators and BS denotes a beam splitter. (b)
Quantum scissors device for the nonlocal generation of arbitrary
VOPS, where $\ket{\rm in}=\ket{\alpha}$ is a coherent input state,
which is truncated to a qubit state $\ket{\rm out}$; $D_i$ are
single-photon photodetectors, and PS denotes a phase shift (0 or
$\pi$), which is applied with a specific probability to decohere
the state $\ket{\rm out}$, i.e., to decrease its coherence factor
$x=\sqrt{p(1-p)}$ to a desired value. } \label{fig_setup}
\end{figure}

In this work, we use photon-number encoding of qubits, and, thus,
we consider Wigner tomographic methods for their reconstruction.
For example, a two-mode state (say $\rho$), which is generated by
mixing a VOPS state with the vacuum at a balanced BS, can be
reconstructed by homodyne tomography by locally mixing each mode
of $\rho$ with a high-intensity classical beam (i.e., a local
oscillator), as shown in Fig.~\ref{fig_setup}(a). The
reconstructed Wigner functions and, thus, also two-mode density
matrices enable the calculation of any quantifiers of two-qubit
correlations and the corresponding potentials for single-mode VOPS
states. The feasibility of this homodyne-QST-based approach has
already been experimentally demonstrated, but only in a special
case of the input single-photon Fock state for testing Bell
nonlocality~\cite{Babichev2004, Fuwa2015} and EPR
steering~\cite{Fuwa2015}.

This paper is organized as follows: The concept of entanglement
potentials is recalled and the potentials for EPR steering and
Bell nonlocality are introduced and analyzed in detail in
Sec.~\ref{Sec_IdealPotentials} assuming ideal experimental
conditions. These concepts are generalized in
Sec.~\ref{Sec_RealPotentials} for realistic experimental
conditions by including the effects of phase damping and
unbalanced beam-splitting (corresponding to amplitude damping).
The phase-space approach to describe nonclassicality using the
Wigner and generalized Wigner functions (i.e., Cahill-Glauber
functions) is given in Sec.~\ref{Sec_Wigner} based on the
definitions summarized in Appendix~\ref{Appendix_Wigner}. Feasible
experimental setups for the generation of the VOPS states and the
tomographic reconstruction of the corresponding two-qubit states
are described in Sec.~\ref{Sec_ExperimentalSetups}. In
Sec.~\ref{Sec_CV} we briefly discuss how nonclassicality
potentials can be defined and applied for higher- or even
infinite-dimensional systems using NC witnesses rather than NC
measures. We conclude in Sec.~\ref{Conclusions}.

\section{Ideal nonclassicality potentials} \label{Sec_IdealPotentials}

Here, by generalizing the idea of entanglement potentials of
Asb\'oth \textit{et al.}~\cite{Asboth2005}, we define other NC
potentials, i.e., those related to quantum steering and Bell
nonlocality and, then, we use them to classify the NC of
single-qubit optical states. We first analyze these potentials
under ideal conditions assuming no damping and a perfectly
balanced BS.

We consider single-qubit optical states defined as (coherent or
incoherent) superpositions of the vacuum $\ket{0}$ and the
single-photon Fock state $\ket{1}$, which are (for brevity)
referred to as VOPS states, and given by a general density matrix,
\begin{equation}
\sigma(p,x) = \sum_{m,n=0}^1\sigma_{mn} \ket{m} \bra{n} =
\left[\begin{array}{cc}
1-p&x\\
x^*&p
\end{array}\right],
 \label{sigma}
\end{equation}
where $p\in[0,1]$ is the probability of measuring a single photon,
$p=\bra{1}\sigma\ket{1}$, and $x$ is a coherence parameter
satisfying $|x|\in[0,\sqrt{p(1-p)}]$. When referring to the VOPS
encoding of qubit states, the only classical state of
$\sigma(p,x)$ is for $p=0$, corresponding to the vacuum.

 \begin{figure}[hbt!]
{\includegraphics[width=\columnwidth]{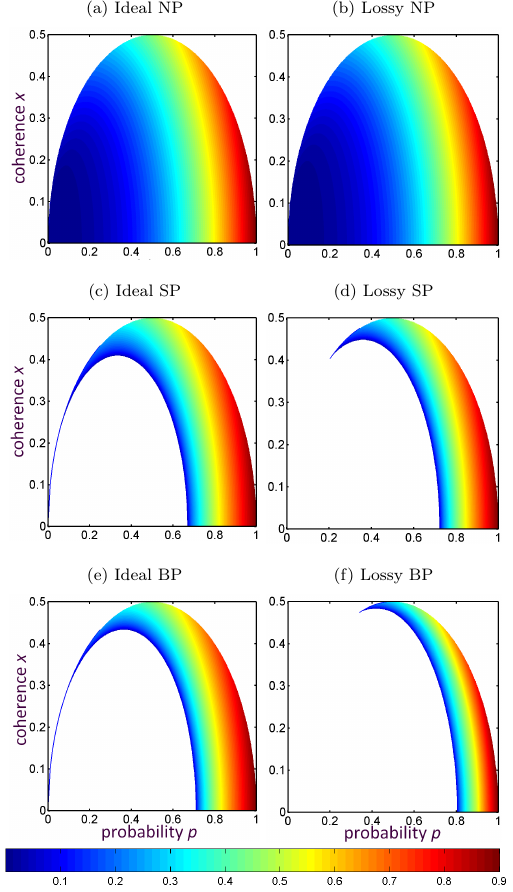}}\hspace*{0pt}\\ %
\caption{Ideal and imperfect nonclassicality potentials for VOPS
states $\sigma(p,x)$ showing the effects of phase damping and
unbalanced BS for: $q=0$, $r=1/\sqrt{2}$ (first column) and
$q=0.1$, $r^2=0.6$ (second column): (a,b) negativity potentials
$\GNP$, (c,d) steering potentials $\GSP$, and (e,f) Bell
nonlocality potentials $\GBP$. It is seen that these imperfections
have the smallest effect on $\GNP$ and the largest effect on
$\GBP$. In panels: (a,b,c,e) the shown potentials for pure states
vanish only for $p=0$, while in (d) $\GSP=0$ for $p\in[0,0.204]$,
and in (f) $\GBP=0$ for $p\in[0,0.339]$.}
 \label{fig_rainbow}
 \end{figure}
 \begin{figure}[hbt!]
\begin{center}
{\includegraphics[width=\columnwidth]{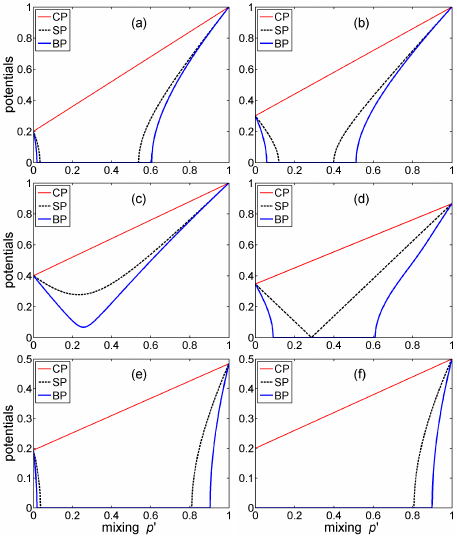}}
\caption{Effect of mixing two single-qubit states on NC
potentials: Ideal (a,b,c) and lossy (d,e,f) NC potentials for
single-qubit states $\sigma'(p,p')$, given in
Eq.~(\ref{sigmaPrime}), versus the mixing parameter $p'$ for
chosen values of the probability $p$ equal to: (a) 0.2, (b) 0.3,
and (c--f) 0.4. Unbalanced BS is assumed only in: (d) for $r=1/2$,
and (e) for $r=1/4$. Phase damping with $q=1/2$ is only assumed in
(f). Graphically, the states $\sigma'(p,p')$ lie on the cross
sections of Figs.~\ref{fig_3regions}(a,b,e) connecting the points
$(p,x)=(1,0)$ and $(p,\sqrt{p(1-p)})$, for fixed values of $p$.}
 \label{fig_mixing}
\end{center}
 \end{figure}
 \begin{figure}[hbt!]
\begin{center}
{\includegraphics[width=\columnwidth]{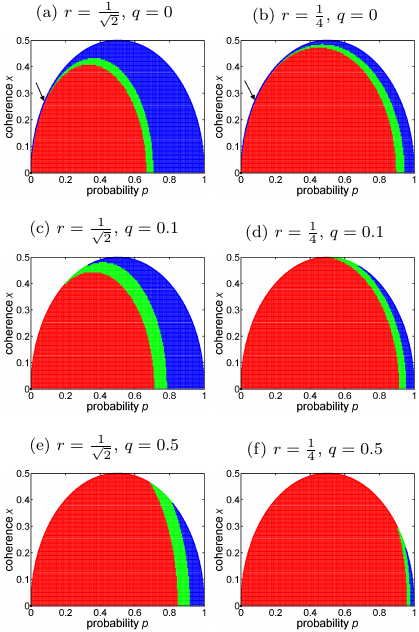}}
\caption{Hierarchy of quantum correlations of VOPS states
$\sigma(p,x)$ versus probability $p=\bra{1}\sigma\ket{1}$ and
coherence parameter $|x|=|\bra{0}\sigma\ket{1}| \in
[0,\sqrt{p(1-p)}]$ for given values of the beam-splitter
reflection ($r$) and dephasing ($q$) parameters. Red, green, and
blue regions mark the states for which it holds: (i)
$\BP=0,\SP=0,\CP>0$; (ii) $\BP=0,\SP>0,\CP>0$; and (iii)
$\BP>0,\SP>0,\CP>0$, respectively. The only VOPS state with
$\BP=\SP=\CP=0$ is for $p=x=0$. The arrows in (a,b) indicate pure
states for which $\BP=\SP=\CP>0$, except for one point. Note that
the transitions between the three regimes occur in panel (a) at
$p=2/3$ and $1/\sqrt{2}$, if $x=0$.}
 \label{fig_3regions}
\end{center}
 \end{figure}
 \begin{figure}[hbt!]
\begin{center}
{\includegraphics[width=\columnwidth]{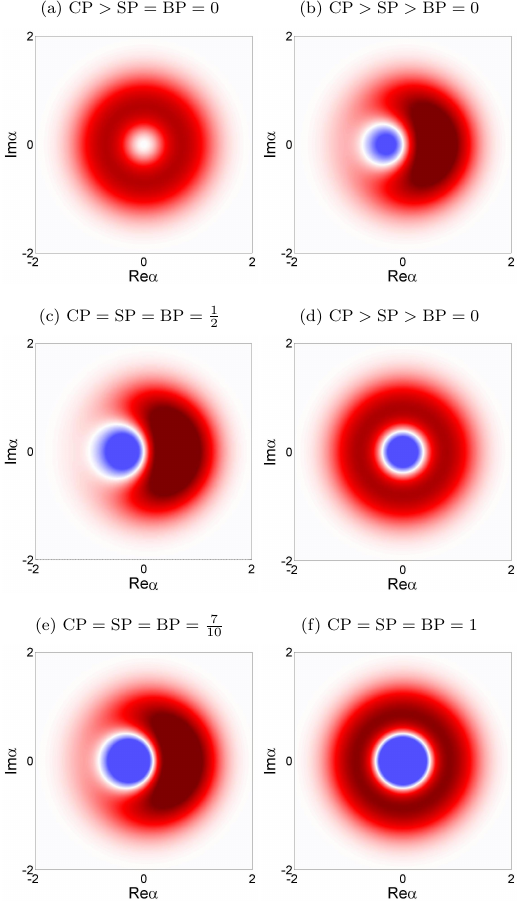}}\hspace*{0pt}\\ %
\caption{Wigner functions $W(\alpha)$ for single-qubit states
$\sigma(p,x)$, for chosen values of the single-photon probability
$p$ and the coherence parameter $x$ showing the hierarchy of
potentials for quantum correlations, which are summarized in
Table~I. Wigner functions for: (a) $\sigma(0.5,0)$, (b)
$\sigma(0.5,0.37)$, (c) $\sigma(0.5,0.5)$, (d) $\sigma(0.7,0)$,
(e) $\sigma[ p,\sqrt{p(1-p)}]$ with $p=0.7$, and (f)
$\sigma(1,0)$. The darker red, the larger positive values of the
Wigner functions; while the darker blue, the more negative values;
white color corresponds to $W(\alpha)=0$. $W(\alpha)$ varies in
the ranges: (a) $[0,0.23]$, (b) $[-0.14,0.50]$, (c)
$[-0.23,0.60]$, (d) $[-0.25,0.25]$, (e) $[-0.39,0.54]$, and (f)
$[-0.64,0.28]$. The negative regions (marked by blue) of the
Wigner functions clearly show the nonclassicality of the
represented states. Note that the state shown in (a) is
nonclassical, although its Wigner function is nonnegative. We did
not show here the trivial case of the Gaussian Wigner function for
the vacuum state when $\CP=\SP=\BP=0$.}
 \label{fig_Wigner}
 \end{center}
 \end{figure}
 \begin{figure}[hbt!]
\begin{center}
{\includegraphics[width=\columnwidth]{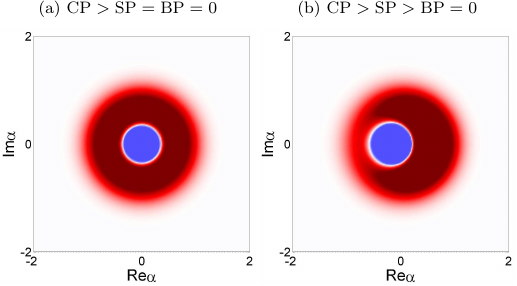}}\hspace*{0pt}\\ %
\caption{Cahill-Glauber function $W^{(1/2)}(\alpha)$ for the
single-qubit states (a) $\sigma(0.5,0)$ and (b) $\sigma(0.5,0.37)$
revealing the hierarchy of NC potentials. The states are the same
as in the corresponding panels (a,b) in Fig.~\ref{fig_Wigner} for
the Wigner function $W^{(0)}(\alpha)$.  $W^{1/2}(\alpha)$ changes
over the ranges: (a) $[-1.27,0.57]$ and (b) $[-1.49,1.17]$, which
correspond, respectively, to the ranges: (a) $[0,0.23]$ and (b)
$[-0.14,0.50]$ for $W^{(0)}(\alpha)$. The negative values of
$W^{(1/2)}(\alpha)$ clearly show the NC character of the states,
even if the corresponding $W^{(0)}(\alpha)$ is nonnegative in the
entire phase space.}
 \label{fig_Cahill}
 \end{center}
 \end{figure}
 \begin{figure}[hbt!]
\begin{center}
{\includegraphics[width=\columnwidth]{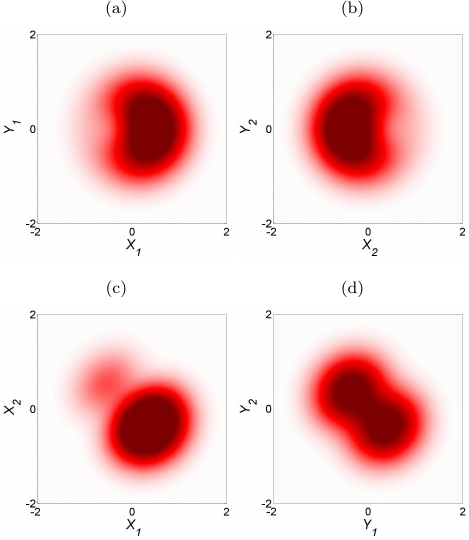}}\hspace*{0pt}\\ %
\caption{Marginal distributions of two-mode Wigner functions
$W(\alpha_1,\alpha_2)$, i.e.: (a) $W(X_1,Y_1)$, (b) $W(X_2,Y_2)$,
(c) $W(X_1,X_2)$, and (d) $W(Y_1,Y_2)$, where $X_i=\RE(\alpha_i)$
and $Y_i=\IM(\alpha_i)$, for single-photon two-mode states
$\rho(p,x)$ assuming $p=0.5$ and $x=0.37$. This state is steerable
in the three-MS, but Bell local (so unsteerable in the two-MS),
and it corresponds to $\sigma(p,x)$ shown in
Fig.~\ref{fig_Wigner}(b). The maximum values of these non-negative
Wigner functions are: (a,b) 0.50, (c) 0.67, and (d) 0.39.}
 \label{fig_marginals}
 \end{center}
 \end{figure}

 \begin{figure*}[hbt!]
\begin{center}
{\includegraphics[width=2\columnwidth]{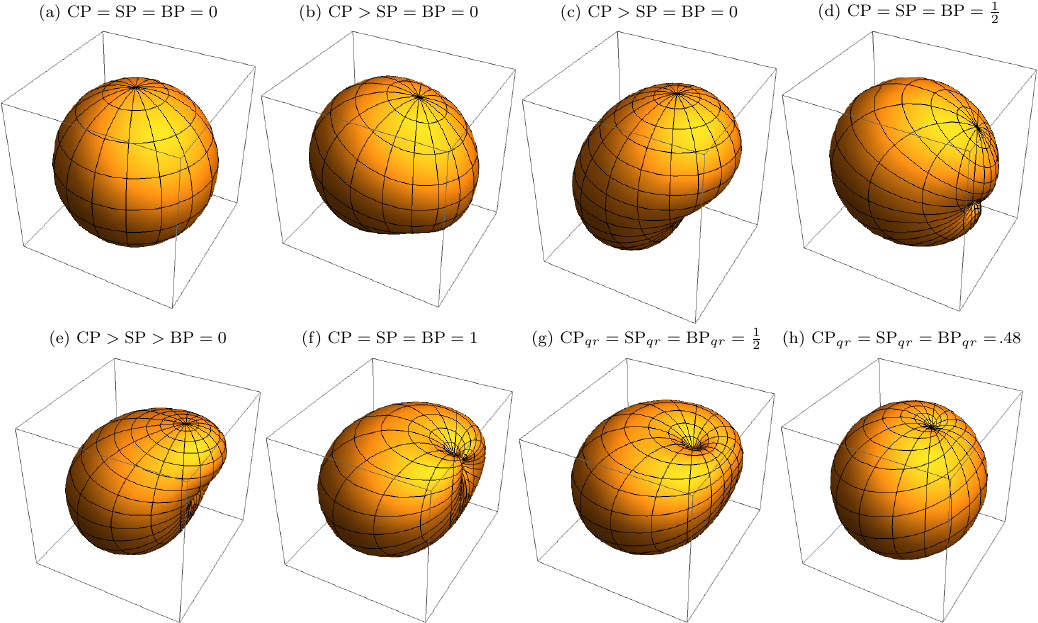}}\hspace*{0pt}\\ %
\caption{Angular-momentum probability surfaces (AMPSs) for chosen
two-mode states $\rho_{qr}(p,x)$, given in Eq.~(\ref{RhoOutGen})
corresponding to a qutrit, revealing the hierarchy of the ideal
(a-f) and lossy (g,h) potentials for entanglement ($\CP$ and
$\GCP$), steering ($\SP$ and $\GSP$), and Bell nonlocality ($\BP$
and $\GBP$), respectively. In the ideal cases the BS is balanced
($r=1/\sqrt{2}$) and  no phase damping occurs ($q=0$); while for
the non-ideal cases we set: (g) phase damping with $q=1/2$ for a
balanced BS; and (h) unbalanced BS with $r=1/4$ and no phase
damping. The chosen two-mode states are: (a)
$\rho(0,0)=\ket{0}\bra{0}$, (b) $\rho(0.1,0)$, (c)
$\rho(\frac12,0)$, (d) $\rho(\frac12,\frac12)$, (e) $\rho(0.7,0)$,
and (f,g,h) $\rho(1,0)=\ket{1}\bra{1}$. In (h), a more precise
value of the three potentials is $0.4841$.}
 \label{fig_AMPS}
 \end{center}
 \end{figure*}
\begin{table*}
\caption{Four regimes of vanishing or nonvanishing two-mode
nonclassicality correlation potentials revealing the hierarchy of
the classes of single-qubit correlations in VOPS states,
$\sigma(p,x)$, depending on the single-photon probability $p$ and
the coherence parameter $x$. Here, $x_{\max}=\sqrt{p(1-p)}$, while
$x_S$ and $x_B$ are given in Eqs.~(\ref{xS}) and (\ref{xB}),
respectively. } \label{table1}
\begin{tabular}{c c c c c c c}
    \hline\hline
     Regime & Entanglement  &  Steering             & Bell nonlocality & Single-photon & Coherence & Examples of states \\
            & potential     & potential in three-MS & potential        & probability   & parameter & shown in figures \\
    \hline
        I   & $\NP=0$       & $\SP=0$ & $\BP=0$     & $p=0$             & $x=0$                   & \ref{fig_AMPS}(a)   \\
        II  & $\NP>0$       & $\SP=0$ & $\BP=0$     & $p\in(0,\frac23]$ & $|x|\in(0,x_S]$         & \ref{fig_Wigner}(a), \ref{fig_Cahill}(a), \ref{fig_AMPS}(b), \ref{fig_AMPS}(c) \\
        III & $\NP>0$       & $\SP>0$ & $\BP=0$     & $p\in(0,\frac23]$ & $|x|\in(x_S,x_B]$       & \ref{fig_Wigner}(b), \ref{fig_Wigner}(d), \ref{fig_Cahill}(b) \\
&&&&\& $p\in(\frac23,\frac1{\sqrt{2}}]$&$|x|\in(0,x_B]$&\ref{fig_AMPS}(e)\\
        IV  & $\NP>0$       & $\SP>0$ & $\BP>0$     & $p\in(0,\frac1{\sqrt{2}}]$ & $|x|\in(x_B,x_{\max}]$  & \ref{fig_Wigner}(c),   \ref{fig_AMPS}(d) \\
        &   &  &                                    & \& $p\in(\frac1{\sqrt{2}},1]$ & $|x|\in(0,x_{\max}]$& \ref{fig_Wigner}(e), \ref{fig_Wigner}(f), \ref{fig_AMPS}(f) \\
    \hline
    \hline
\end{tabular}
\end{table*}
\subsection{Entanglement potentials for a single qubit}

According to the approach of Ref.~\cite{Asboth2005}, a NC measure
of a single optical qubit $\sigma$ can be defined by the
entanglement of the output state $\rho_{\rm out}$ of an auxiliary
lossless balanced beam-splitter (BS) with the state $\sigma$ and
the vacuum $\ket{0}$ at the inputs [see Fig.~\ref{fig_setup}(a)],
i.e.,
\begin{equation}
  \rho_{\rm out} = U_{\rm BS} (\sigma\otimes \ket{0}\bra{0} )U^\dagger_{\rm
  BS},
\label{RhoOut1}
\end{equation}
given in terms of the unitary transformation $U_{\rm
BS}=\exp(-iH\theta)$; with the Hamiltonian
$H=\frac12i(a_{1}^\dagger a_{2}-a_{1} a_{2}^\dagger)$, where
$a_{1,2}$ ($a_{1,2}^\dagger$) are the annihilation (creation)
operators of the input modes and, for simplicity, we set
$\hbar=1$.  Moreover, the BS parameter $\theta$ defines the
reflection and transmission coefficients, as $r=\sin(\theta/2)$
and $t=\cos(\theta/2)$, respectively.

First, we set $\theta=\pi/2$ for a balanced BS. Linear
transformations (as discussed in greater detail in
Sec.~\ref{Sec_Wigner}) do not change the global NC of an optical
field. Thus, the output state $\rho_{\rm out}$ is entangled if and
only if the input state $\sigma$ is nonclassical. Let us recall
that coherent states (so infinite-dimensional states except the
vacuum) and their statistical mixtures are the only classical
states. Thus, an arbitrary finite-dimensional single-mode optical
state (except the vacuum) is nonclassical and if it is mixed with
the vacuum on a BS, then the output two-mode state is entangled.

In a special case, by mixing an arbitrary single-qubit optical
state $\sigma(p,x)$ with the vacuum on a perfect balanced BS and
assuming no phase and amplitude dissipation, the following state
$\rho_{\rm out}\equiv \rho(p,x)$ is generated:
\begin{equation}
  \rho(p,x) =\left[\begin{array}{cccc}
1-p& -\frac1{\sqrt{2}}x &\frac1{\sqrt{2}}x& 0\\
-\frac1{\sqrt{2}}x^*&\frac12 p&-\frac12p&0\\
\frac1{\sqrt{2}}x^*& -\frac12 p&\frac12p&0\\
0&0&0&0\\
\end{array}\right].
\label{RhoOut}
\end{equation}
To quantify the NC of $\sigma$, we consider the entanglement
potential~\cite{Asboth2005}:
\begin{eqnarray}
  \CP(\sigma) &=& E (\rho_{\rm out}), \label{EP1}
\end{eqnarray}
which is defined by, e.g., the Wootters
concurrence~\cite{Wootters1998}:
\begin{equation}
  C({\rho})=\Theta \Big(2\max_j\lambda_j-\sum_j\lambda_j\Big),
  \label{concurrence}
\end{equation}
given in terms $\Theta(x)\equiv\max(x,0)$, $\lambda^2 _{j} =
\mathrm{eig}[{\rho_{\rm out} } (\sigma_2\otimes \sigma_2)\rho_{\rm
out}^{\ast }(\sigma_2\otimes \sigma_2)]_j$, $\sigma_2$ is a Pauli
operator, and asterisk denotes complex conjugation. The
concurrence is a monotone of the entanglement of
formation~\cite{HorodeckiReview}. As shown in~\cite{Adam2015a},
the concurrence of $\rho(p,x)$ is given simply by the
single-photon probability $p$ and the coherence parameter $x$,
\begin{eqnarray}
  \CP[\sigma(p,x)] &=& E [\rho(p,x)] = p. \label{EP2}
\end{eqnarray}
Thus, a single-qubit state $\sigma(p,x)$ has a nonzero
entanglement potential, $\CP[\sigma(p,x)]>0$ for any $p>0$ and
$|x|\in [0,\sqrt{p(1-p)}]$, i.e., for all allowed values of the
parameters except $p=x=0$.

In addition to the concurrence, one can apply the negativity,
which is arguably the most popular measure of entanglement. The
negativity for two qubits in a state $\rho_{\rm out}$ is defined
by~\cite{HorodeckiReview}:
\begin{equation}
  N({\rho_{\rm out}})=\max\big[0,-2\min{\rm eig}(\rho_{\rm
out}^{\Gamma})\big],
  \label{negativity}
\end{equation}
where $\rho_{\rm out}^{\Gamma}$ is the partial transpose of
$\rho_{\rm out}$ with respect to either qubit. Thus, the
negativity potential (NP) of a single-qubit state $\sigma$ is
defined as the negativity $N$ of the two-qubit output state
$\rho_{\rm out}$, i.e.,
\begin{eqnarray}
  \NP(\sigma) = N(\rho_{\rm out}). \label{NP}
\end{eqnarray}
The explicit formula for $\NP$ for an arbitrary single-qubit state
$\sigma(p,x)$ reads~\cite{Adam2015a}:
\begin{equation}
  \NP[\sigma(p,x)]=\frac{1}{3} \left[2 {\rm Re}\left(\sqrt[3]{2 \sqrt{a_1}+2 a_2}\right)+p-2\right],
 \label{NPgeneral}
\end{equation}
where
\begin{eqnarray}
a_{1} &=& a_2^2-2 \big[5 (p-1) p+6 |x|^2+2\big]^3,
\nonumber \\
a_{2} &=& 14 p^3-21 p^2+15 p+9 (p-2) |x|^2-4, \label{NPgeneral2}
\end{eqnarray}
which depends, in general, on the absolute value of the coherence
parameter, $|x|$, which is not the case for the concurrence
potential. Note that $\NP[\sigma(p,x)]>0$ iff
$\CP[\sigma(p,x)]>0$, because the negativity and concurrence are
good measures of two-qubit entanglement. For some classes of
states, including pure states and Werner states,
$\NP[\sigma(p,x)]$ and $\CP[\sigma(p,x)]$ are the same, although
they are different in general.

An entanglement potential can also be defined via the relative
entropy of entanglement (REE)~\cite{HorodeckiReview}, as studied
in, e.g., Refs.~\cite{Asboth2005,Adam2015a,Adam2015b}.
Unfortunately, no analytical formulas are known for the REE of
$\rho(p,x)$ assuming general parameters $p$ and $x$, so here we
limit our study of entanglement potentials to the $\CP$ and $\NP$.

\subsection{Ideal steering potentials for a single qubit}

EPR steering is a type of quantum nonlocality between two parties
(qubits or modes) that is in general distinct from both
entanglement and Bell nonlocality. In the original meaning, it
describes the ability of one observer to influence another party's
(qubit's) state via local measurements on its system (qubit).
According to Ref.~\cite{Wiseman2007}, EPR steering arises from the
quantum correlations exhibited by quantum systems, enabling the
verification of entanglement even when complete characterization
of one of the subsystems is lacking. Thus, EPR steering can be
interpreted as a stronger form of entanglement such that can be
detected even by untrusted detectors in one subsystem.
Specifically, by considering the setup shown in
Fig.~\ref{fig_setup}(a), this interpretation could correspond to
assuming low- (high-) quality detectors used by, e.g., Alice (Bob)
for their homodyne QST. Inspired by this interpretation,
applications of quantum steering have been found for quantum
cryptography~\cite{UolaReview} and enhanced
metrology~\cite{Yadin2021,Lee2023}. Moreover,
temporal~\cite{Chen2014,Bartkiewicz2016exp} and
spatiotemporal~\cite{Chen2017} analogues of standard (spatial)
quantum steering have also been found and applied in quantum
cryptography~\cite{Bartkiewicz2016crypto}, as well as for
quantifying non-Markovianity~\cite{Chen2016}, or witnessing
quantum scrambling~\cite{Lin2021} and nonclassical correlations in
quantum networks~\cite{Chen2017}.

Here, inspired by Ref.~\cite{Asboth2005}  entanglement potential
for a single optical mode defined via a two-mode entanglement
measure, we propose to define a steering potential for a single
optical qubit (or mode) quantified by a measure of standard
two-qubit (or two-mode) EPR steering. In the following, for
simplicity, we consider the standard Costa-Angelo measure of
steering~\cite{Costa2016} for which an analytical formula can be
found. Of course, other measures of steering can also be used in
defining steering potentials, including the steerable
weight~\cite{Skrzypczyk2014} and/or the steering
robustness~\cite{Piani2015}; however, such definitions would be
based  on numerical calculations  for general single-qubit states
except some simple classes of states.

The steering potential quantified by the Costa-Angelo measure of
steering~\cite{Costa2016} in a three-measurement scenario
(three-MS), corresponding to measuring the three Pauli operators
on qubits of both parties, can be defined as
\begin{eqnarray}
  \SP'(\sigma)=S_{\rm CA}^{(3)}(\rho)= \frac{\Theta(\sqrt{\tr R}-1)}{\sqrt{3}-1},\label{SP3}
\end{eqnarray}
given in terms of the correlation function $R=T^TT$, where the
elements of the matrix $T$ are the two-qubit Stokes parameters,
$T_{ij}=\mathrm{Tr}[\rho(\sigma_{i}\otimes\sigma_{j})]$. Note that
the correlation matrix $R$, and thus $S_{\rm CA}^{(3)}$, can be
determined even experimentally without full QST, as recently
demonstrated in~\cite{Abo2023}. To show this explicitly, we recall
the Bloch representation of a general two-qubit state $\rho$:
\begin{equation}
\rho  =  \frac{1}{4}\Big(I\otimes
I+\boldsymbol{u}\cdot\boldsymbol{\sigma}\otimes
I+I\otimes\boldsymbol{v}\cdot\boldsymbol{\sigma}+\!\!\!\sum
\limits_{n,m=1}^{3}T_{nm}\,\sigma _{n}\otimes \sigma _{m}\Big),
\label{rhoGeneral}
\end{equation}
where $\boldsymbol{\sigma}=[\sigma_{1},\sigma_{2},\sigma_{3}]$ are
the Pauli matrices.  Moreover in Eq.~(\ref{rhoGeneral}), the
elements of the Bloch vectors $\boldsymbol{u}=[u_{1},u_{2},u_{3}]$
and $\boldsymbol{v}=[v_{1},v_{2},v_{3}]$ are
$u_{i}=\mathrm{Tr[}\rho(\sigma_{i}\otimes I)]$ and
$v_{i}=\mathrm{Tr[}\rho(I\otimes\sigma_{i})]$, respectively, and
$I$ is the single-qubit identity operator. Thus, the
reconstruction of the correlation matrix $R=T^TT$ of $\rho$,
without reconstructing the Bloch vectors $\boldsymbol{u}$ and
$\boldsymbol{v}$, enables the calculation of the steering and
nonlocality measures and, thus, the corresponding potentials
discussed below.

The steering potential can be defined in a modified way:
\begin{equation}
  \SP(\sigma)= S^{(3)}(\rho) = \sqrt{\tfrac 12\Theta(\tr R-1)},
  \label{SPprime}
\end{equation}
which corresponds to the three-MS steering measure applied in
Refs.~\cite{Fan2021,Yang2021,Abo2023}). Note that $S_{\rm
CA}^{(3)}$ and $S^{(3)}$ are both measures of the violation of the
steering inequality derived by Cavalcanti, Jones, Wiseman, and
Reid (CJWR) in the three-MS~\cite{Cavalcanti2009}. The two
steering potentials are monotonically related for arbitrary
single-qubit states $\sigma$ by
\begin{equation}
  \SP'(\sigma)=\frac{\sqrt{2[\SP(\sigma)]^2+1}-1}{\sqrt{3}-1}\le \SP(\sigma).
  \label{SteeringRelation}
\end{equation}
in analogy to the corresponding relation for the steering
measures~\cite{Abo2023}. In this paper we focus on $\SP(\sigma)$
because it reduces to the entanglement potential for any two-qubit
pure states, $\sigma[p,\sqrt{p(1-p)}]\equiv \ket{\psi}\bra{\psi}$.
On the other hand, $\SP'(\sigma)$ calculated for experimentally
reconstructed states gives usually a better agreement with
theoretical predictions (see Ref.~\cite{Abo2023} for comparison of
experimentally determined $S_{\rm CA}^{(3)}$ and $S^{(3)}$ for
Werner-like states). Thus, we present both definitions.

We find that the steering potential in the three-MS for a
single-qubit state $\sigma(p,x)$ is given by
\begin{equation}
  \SP(\sigma)=\sqrt{\Theta(3p^2-2p+2|x|^2)},
  \label{SP_out}
\end{equation}
clearly depending on the coherence parameter $|x|$, which is not
the case for $\CP(\sigma)$. Thus, we find that a given state
$\sigma(p,x)$ has a nonzero steering potential,
$\SP[\sigma(p,x)]>0$ for: (i) $p\in (0,2/3]$ if $|x|\in (
x_S(p),\sqrt{p(1-p)}]$, where
\begin{equation}
  x_S(p)=\sqrt{p(1-3p/2)},
  \label{xS}
\end{equation}
and (ii) $p\in (2/3,1]$ if $|x|\in [0,\sqrt{p(1-p)}]$.

To explain a rapid decrease and vanishing of the SP by introducing
even a slight decoherence of a pure state
$\sigma[p,\sqrt{p(1-p)}]$ for small $p$, as seen in
Figs.~\ref{fig_rainbow}(c), \ref{fig_mixing}(a), and
\ref{fig_mixing}(b), we introduce a decoherence factor
$\kappa\in[0,1]$, such that $x=\kappa\sqrt{p(1-p)}$. By analyzing
Eq.~(\ref{SP_out}), one readily finds that  the decoherence factor
should satisfy
\begin{equation}
  \kappa>\kappa_0=\frac{2-3p}{2-2p},
  \label{kappa}
\end{equation}
to guarantee that $\SP(\sigma)>0$. Thus, we see that the steering
potential is nonzero for any value of $\kappa$ (and so $x$) if
$p>2/3$. However, if $p=0.1$ (0.2), $\SP(\sigma)>0$ for
$\kappa>0.94$ ($>0.875$). This clearly explains a rapid
disappearance of the steering potential shown by the thin curve on
the left-hand side of Fig.~\ref{fig_rainbow}(c). Moreover, even a
more rapid loss of the nonlocality potential can be seen in
Fig.~\ref{fig_rainbow}(e), because a vanishing $\SP$ implies a
vanishing $\BP$.

\subsection{Bell nonlocality potentials for a single qubit}

Single-qubit Bell nonlocality potentials can be introduced via
Bell nonlocality measures of a given two-qubit state $\rho$
quantifying the violation of the Bell inequality in the
Clauser-Horne-Shimony-Holt form (denoted as Bell-CHSH)
\cite{Clauser1969}:
\begin{equation}
|\langle {\cal
B}\rangle_{\rho}|\equiv\big|\big\langle\boldsymbol{a}\cdot
\boldsymbol{\sigma }\otimes (\boldsymbol{
b}+\boldsymbol{b}^{\prime })\cdot \boldsymbol{\sigma
}+\boldsymbol{a}^{\prime }\cdot \boldsymbol{\sigma }\otimes
(\boldsymbol{b}-\boldsymbol{b}^{\prime })\cdot \boldsymbol{\sigma
}\big\rangle_{\rho}\big|\leq 2, \label{CHSH}
\end{equation}
given in terms of the Bell-CHSH operator ${\cal B}$, where
$\boldsymbol{a}, \boldsymbol{a'}, \boldsymbol{b},
\boldsymbol{b'}\in\mathbb{R}^3$ are unit vectors describing
measurement settings. As described by Horodecki \emph{\textit{et
al.}}~\cite{Horodecki1995}, the maximum possible violation of the
Bell-CHSH inequality in Eq.~(\ref{CHSH}) considered over all
measurement settings, can be used as a Bell nonlocality measure,
i.e., $\max_{\nu}\langle {\cal B}\rangle_{\rho}=2\sqrt{{\cal
M}(\rho)}$, where ${\cal M}(\rho)$ is the sum of the two largest
eigenvalues of the correlation matrix $R(\rho)$. The Bell-CHSH
inequality is satisfied if and only if ${\cal M}(\rho)\le 1$. To
make our comparison of various types of quantum correlations
consistent, as based on measures and potentials defined in the
range [0,1], the Bell nonlocality measure of
Ref.~\cite{Horodecki1995} is often rescaled as $B(\rho) =
\sqrt{\Theta[{\cal M}(\rho)-1]}$ (see, e.g.,
Refs.~\cite{Adam2004,Karol2017,Yang2021}). Thus, we define a Bell
nonlocality potential as
\begin{eqnarray}
  \BP(\sigma) &=&  B(\rho) = \sqrt{\Theta[{\cal M} (\rho)-1]} \nonumber \\
  &=& \sqrt{\Theta\big\{\tr R -\min[{\rm eig}(R)]-1\big\}}.
\label{nonlocalityB}
\end{eqnarray}
This measure is monotonically related to the Costa-Angelo measure
of steering~\cite{Costa2016} defined in the two-measurement
scenario (two-MS), which corresponds to measuring two Pauli
operators on qubits of both parties~\cite{Costa2016}.
Specifically, the related nonlocality potential reads
\begin{eqnarray}
  \BP'(\sigma) = S_{\rm CA}^{(2)}(\rho) = \frac{\Theta\big\{\sqrt{\tr R-\min[{\rm
  eig}(R)]}-1\big\}}{\sqrt{2}-1},
\label{S2}
\end{eqnarray}
which is simply related to $\BP(\sigma)$ as
\begin{equation}
  \BP'(\sigma)=\frac{\sqrt{[\BP(\sigma)]^2+1}-1}{\sqrt{2}-1} \le \BP(\sigma).
  \label{BvsB}
\end{equation}
Note that $B,B',\BP,\BP'\in[0,1]$ for arbitrary two-qubit states.
We find that
\begin{eqnarray}
  \min[{\rm eig}(R)]&=& \frac12\Big( 1+p(5p-4)+4|x|^2
    \nonumber \\
  &&-(1-p)\sqrt{(1-3p)^2+8|x|^2}\Big),
  \nonumber \\
  \tr R &=& 1 - 4 p + 6 p^2 + 4 x^2,
\label{N1}
\end{eqnarray}
so the Bell nonlocality potential $\BP$ for a general state
$\sigma(p,x)$ with $p\in[0,1]$ and $x\in[0,\sqrt{p(1-p)}]$ becomes
\begin{eqnarray}
  \BP[\sigma(p,x)]&=&\Big\{\Theta\Big[\frac{1}{2} \big(7 p^2+(1-p)
  \sqrt{(1-3 p)^2+8 |x|^2}\nonumber \\
  &&-4 p+4 |x|^2-1\big)\Big]\Big\}^{1/2}.
  \label{N}
\end{eqnarray}
We find that a given state $\sigma(p,x)$ has a nonzero nonlocality
potential, $\BP[\sigma(p,x)]>0$ for: (i)
$p\in(0,\frac1{\sqrt{2}}]$ with $|x|\in (x_B(p),\sqrt{p(1-p)}]$,
where
\begin{equation}
  x_B(p)= \frac1{\sqrt{2}}\sqrt{1+p-3 p^2-(1-p)\sqrt{1-p^2}},
  \label{xB}
\end{equation}
and (ii) $p\in(\frac1{\sqrt{2}},1]$ with $|x|\in
[0,\sqrt{p(1-p)}]$.

\subsection{Hierarchy of nonclassicality potentials}

Single-qubit correlations, as quantified by the NC potentials,
satisfy the following hierarchy:
\begin{eqnarray}
  \BP(\sigma) \le \SP(\sigma) \le \NP(\sigma) \le \CP (\sigma),
\label{ideal_hierarchy}
\end{eqnarray}
for an arbitrary state $\sigma(p,x)$. This hierarchy is in close
analogy to that for the corresponding two-qubit correlation
measures (see, e.g.,~\cite{Abo2023}). For single-qubit pure states
$\sigma(p,\sqrt{p(1-p)})=\ket{\psi}\bra{\psi}$, the potentials
become the same,
\begin{equation}
    \BP(\ket{\psi}) = \SP(\ket{\psi}) = \NP(\ket{\psi}) = \CP (\ket{\psi}) =
    p.
  \label{pure_hierarchy}
\end{equation}
Figure~\ref{fig_3regions}(a) shows the hierarchy of ideal NC
potentials, i.e., assuming a lossless and balanced BS. In addition
to the vacuum (for $p=x=0$), which is the only separable VOPS
state, this hierarchy includes the states with the potentials for:
(i) non-steerable entangled states (corresponding to the red
region), (ii) steerable states but Bell local (in the green
region), and (iii) Bell nonlocal states (in the blue region). We
can see in this figure that a VOPS state has nonvanishing SP or BP
if either $p$ is sufficiently large (and then independent of $x$)
or if smaller values of $p$ are accompanied by a sufficiently
large $x$. The same conclusion can be drawn by analyzing
Figs.~\ref{fig_rainbow}(c,e) and Eqs.~(\ref{xS}), (\ref{kappa}),
and (\ref{xB}).

Figures~\ref{fig_mixing}(a,b,c) show the ideal NC potentials and
their hierarchy for the mixed states defined as
\begin{equation}
  \sigma'(p,p') = p'\ket{1}\bra{1}+(1-p')\ket{\psi_p}\bra{\psi_p},
\label{sigmaPrime}
\end{equation}
where $\ket{\psi_p}=\sqrt{p}\ket{1}+\sqrt{1-p}\ket{0}$. These
states lie on the cross sections of some graphs in
Fig.~\ref{fig_3regions} as explained in detail in the caption of
Fig.~\ref{fig_mixing}. In particular, a very narrow region, which
close to $p'=0$ with nonzero $\BP$ and $\SP$, is shown in
Fig.~\ref{fig_mixing}(a) for $p=0.2$.

\section{Realistic nonclassicality potentials} \label{Sec_RealPotentials}

We stress that the standard entanglement potentials of
Ref.~\cite{Asboth2005} are based solely on the special case of
$\rho_{\rm out}$ for a balanced (50/50) BS assuming no
dissipation. Now we analyze how experimental imperfections can
affect the  two-qubit states generated from single-qubit states
given in Eq.~(\ref{sigma}).

We first consider the effect of phase damping. Specifically, the
Kraus operators for a single-qubit phase-damping channel (PDC)
read~\cite{NielsenBook}:
\begin{equation}
E_{0}(q_{i})=|0\rangle\langle0|+\sqrt{1-q_{i}}|1\rangle\langle1|,\quad
E_{1}(q_{i})=\sqrt{q_{i}}|1\rangle\langle1|,\label{Kraus_pdc}
\end{equation}
where $q_i$ (with $i=1,2$) are phase-damping coefficients (rates),
and the Kraus operators satisfy the normalization relation
$\sum_{n=0,1}E_{n}^{\dagger}(q_i)E_{n}(q_i)=I$. Two-qubit phase
damping transforms a given two-mode state $\rho_{{\rm in}}$ to
\begin{equation}
\rho_{{\rm PDC}}=\sum_{i,j}[E_{i}(q_{1})\otimes
E_{j}(q_{2})]\rho_{{\rm in}}[E_{i}^{\dagger}(q_{1})\otimes
E_{j}^{\dagger}(q_{2})]. \label{Kraus_pdc2}
\end{equation}
For simplicity, we analyze the same phase damping rate in both
qubits, so we set $q\equiv q_1=q_2$.

We also consider the effect of an unbalanced BS on the generation
of two-mode states, as given by Eq.~(\ref{RhoOut1}) for $r\neq
t=\sqrt{1-r^2}$. By the inclusion of these effects, we find that
the output state $\rho_{\rm out}$, given in Eq.~(\ref{RhoOut1}),
now generalizes to
\begin{equation}
  \rho_{qr}(p,x) = \left[
\begin{array}{cccc}
 1-p & -Q r x & Q t x & 0 \\
 -Q r x^* & p r^2 & -p Q^2 r t & 0 \\
 Q t x^* & -p Q^2 r t & p t^2 & 0 \\
 0 & 0 & 0 & 0 \\
\end{array}
\right],
  \label{RhoOutGen}
\end{equation}
where $Q = \sqrt{1 - q}$ for the phase damping parameter $q$.
Equation~(\ref{RhoOutGen}) reduces to Eq.~(\ref{RhoOut}) for
$r=t=1/\sqrt{2}$ and $q=0$.

Thus, by considering these imperfections, we can analyze the
entanglement, steering, and Bell nonlocality generalized
potentials corresponding to more realistic experimental
situations, as defined, respectively, by
\begin{eqnarray}
  \GCP(\sigma) &=& C(\rho_{qr}), \label{GCP} \\
  \GNP(\sigma) &=& N(\rho_{qr}), \label{GNP} \\
  \GSP(\sigma) &=& S^{(3)}(\rho_{qr}), \label{GSP}  \\
  \GBP(\sigma) &=& B(\rho_{qr}), \label{GBP}
\end{eqnarray}
and analogously to the related potentials based on $S_{\rm
CA}^{(3)}(\rho_{qr})$ and $S_{\rm CA}^{(2)}(\rho_{qr})$. The
hierarchy relations, given in Eq.~(\ref{ideal_hierarchy}) for the
ideal potentials, simply generalize for the realistic (i.e.,
lossy) NC potentials to
\begin{eqnarray}
  \GBP(\sigma) \le \GSP(\sigma) \le \GNP(\sigma) \le \GCP
  (\sigma).
\label{realistic_hierarchy}
\end{eqnarray}

We find that the concurrence generalized potential reads
\begin{equation}
  \GCP[\sigma(p,x)]=C[\rho_{qr}(p,x)]=2p(1-q)rt=p(1-q)\sin\theta,
  \label{GCP1}
\end{equation}
where the BS parameter $\theta$ is defined below
Eq.~(\ref{RhoOut1}). One can define other entanglement generalized
potentials based on, e.g., the universal witness of entanglement
(UWE) can be defined by $\det\rho^\Gamma$~\cite{Augusiak2008} or
$\Theta(-\det\rho^\Gamma)$ to be consistent with the definitions
of other nonclassicality quantifiers applied in this paper. Note
that an effective experimental method for measuring the UWE
without full QST was described in~\cite{Karol2015b} (although the
method has not been implemented experimentally yet). The
measurement of the concurrence of a two-qubit state requires
usually its full QST. The UWE and the corresponding entanglement
generalized potential ${\rm UWEP}_{qr}$ for the $\theta$-dependent
BS output state reads \begin{equation}
  {\rm UWEP}_{qr}\equiv \Theta[-\det\rho_{qr}^\Gamma(p,x)]=(\tfrac12 p\sin\theta)^4(1-q)^2,
  \label{UWEP}
\end{equation}
being independent of $x$, which is the same as for the concurrence
potential $\GCP$, but contrary to the negativity potential $\GNP$.
Anyway, it holds
\begin{equation}
  {\rm UWEP}_{qr}(\sigma)>0 \Leftrightarrow \GCP(\sigma)>0
\Leftrightarrow \GNP(\sigma)>0,
  \label{iff}
\end{equation}
for any $x$. Note that the analytical expression for $\GNP$, which
generalizes Eq.~(\ref{NPgeneral}), is quite lengthy, so it is not
presented here.

We stress that all the nonclassicality measures and quantifiers
considered in this paper, are independent of the phase of $x$,
although the $R$ matrix depends. So, to have compact formulas for
the $R$ matrix, let us set in the following equations that the
coherence parameter \emph{$x$ is real}. Then the correlation
matrix $R$ reads \begin{equation}
  R= \left[
\begin{array}{ccc}
 4 Z p^2+4 Q^2 r^2 x^2 & 0 &  Y \\
 0 & 4 p^2 Z & 0 \\
 Y & 0 & (1-2 p)^2+4 Q^2 t^2 x^2 \\
\end{array}
\right],
  \label{R_gen}
\end{equation}
where
\begin{eqnarray}
  Y &=& 2 Q r \left(-2 p Q^2 t^2+2 p-1\right) x,\nonumber \\
  Z &=& Q^4 r^2 t^2 =\tfrac14 (1-q)^2\sin^2\theta.
 \label{YZ}
\end{eqnarray}
The eigenvalues of $R$ are found as:
\begin{eqnarray}
  e_{1,2}&=&\frac{1}{2} \left(1+4 \left[ Q^2 x^2+p^2
  (Z+1)-p\right]\pm \sqrt{f_e}\right),\nonumber\\
  e_3 &=& 4p^2 Z=[p(1-q)\sin\theta]^2,
\label{Reigs_gen}
\end{eqnarray}
where
\begin{equation}
  f_e =   \left(\tr R-4 p^2 Z\right)^2-16 Z \left(2
   p^2+2 x^2-p\right)^2.
  \label{fe}
\end{equation}
Thus, we have
\begin{eqnarray}
  \tr R=\sum e_i &=& 4 p (2 p Z+p-1)+4 Q^2 x^2+1,\label{trR_gen}\\
  \min[{\rm eig}(R)]&=&\min(e_2,e_3),
  \label{Reigs_gen2}
\end{eqnarray}
which enable the calculation of the generalized potentials for
$\GSP$ and $\GBP$.

The hierarchy of the lossy NC potentials is plotted in
Figs.~\ref{fig_3regions}(b-f) in comparison to the ideal NC
potentials shown in Fig.~\ref{fig_3regions}(a). The red, green,
and blue regions show, respectively, the regimes II, III, and IV
listed in Table~I; while the point $(p,x)=(0,0)$ indicates the
only separable VOPS state (i.e., the vacuum), which belongs to the
regime~I. It is clearly seen that dephasing and unbalanced beam
splitting considerably decrease the regions of the nonvanishing
steering and nonlocality potentials.

\section{Phase-space and angular-momentum descriptions
of nonclassicality}\label{Sec_Wigner}

\subsection{Wigner and Cahill-Glauber quasiprobability distributions}

To visualize the nonclassicality of the analyzed single- and
two-mode states, we here apply the standard Wigner functions and
their generalizations.

It is well known that linear transformations (including that of a
BS) do not change the global  nonclassicality of states. This can
be convincingly demonstrated using the Cahill-Glauber
$s$-parametrized quasiprobability distribution (QPD), which is
defined for any $s\in[-1,1]$ in Appendix~\ref{Appendix_Wigner}.
Note that the $s$-parametrized QPD reduces in special cases to the
standard Wigner ($W={\cal W}^{(0)}$) and Husimi ($W={\cal
W}^{(-1)}$) functions, which can be measured experimentally, and
to the Glauber-Sudarshan function ($P={\cal W}^{(1)}$), which is
used in the definition of the nonclassicality of optical fields,
but usually cannot be measured experimentally, because of its
singularity (except very special nonclassical fields).

For example, a perfect BS transformation, which is given by the
unitary transformation $U_{\rm BS}$, given below
Eq.~(\ref{RhoOut1}), of an arbitrary input state $\rho_{\rm in}$
(of any dimension) resulting in the two-mode output state
$\rho_{\rm out}$, can equivalently be described by the evolution
in a two-mode phase space of the corresponding QPD given
by~\cite{LeonhardtBook}
\begin{equation}
  {\cal W}_{\rm out}^{(s)}(\alpha_1,\alpha_{2})
  = {\cal W}_{\rm in}^{(s)} (t\alpha_1+r\alpha_2,r\alpha_1-t\alpha_2).
  \label{QPD_BS1}
\end{equation}
This equation implies that the initial QPD is displaced, without
changing its form, along a trajectory in the phase space spanned
by the canonical position ($X_i\equiv{\rm Re}\,\alpha_i$ for
$i=1,2$) and ($Y_i\equiv{\rm Im}\,\alpha_i$) momentum operators.
The trajectory is given by the solution of the corresponding
classical equations of motion. Thus, the global nonclassicality of
the state is unchanged during this evolution.

In this paper we are mainly interested in a special case of the BS
transformation assuming a VOPS state $\sigma$ in one input ports
and the vacuum in another port. Then, Eq.~(\ref{QPD_BS1}) reduces
to
\begin{equation}
  {\cal W}_{\rm out}^{(s)}(\alpha_1,\alpha_{2})
  = {\cal W}_{\rm vops}^{(s)}(t\alpha_1+r\alpha_2)\,
  {\cal W}^{(s)}_{\rm vac}(r\alpha_1-t\alpha_2),
  \label{QPD_BS2}
\end{equation}
where ${\cal W}_{\rm vac}^{(s)}$ is the single-mode-vacuum QPD
given by
\begin{equation}
  {\cal W}_{\rm vac}^{(s)}(\alpha)=\frac1{\pi}T^{(s)}_{00}(\alpha)=
  \frac{2}{\pi(1- s)} \exp\left(-\frac{2}{1- s} |\alpha|^2\right),
\label{QPD_vac}
\end{equation}
and ${\cal W}_{\rm vops}^{(s)}(\alpha)$ is the QPD for an
arbitrary single-mode state $\sigma(p,x)$:
\begin{eqnarray}
  {\cal W}_{\rm vops}^{(s)}(\alpha) &=&
  \frac{1}{\pi}\Big[(1-p) T^{(s)}_{00}(\alpha) + p T^{(s)}_{11}(\alpha) \nonumber\\
  &&\quad + x T^{(s)}_{10}(\alpha) + x^* T^{(s)}_{01}(\alpha)
  \Big],
\label{QPD_VOPS}
\end{eqnarray}
with the functions $T^{(s)}_{nm}(\alpha)$ given explicitly in
Eq.~(\ref{Tnm_VOPS}). Note that Eq.~(\ref{QPD_vac}) is a special
case of Eq.~(\ref{QPD_VOPS}). Another important special case of
that formula is the QPD ${\cal W}_{\rm
1ph}^{(s)}(\alpha)=T^{(s)}_{11} (\alpha)/\pi$ for the
single-photon Fock state:
\begin{equation}
  {\cal W}_{\rm 1ph}^{(s)}(\alpha)=
  \frac{2(4 |\alpha|^2+s^2-1)}{\pi(1- s)^{3}} \exp\left(-\frac{2}{1- s} |\alpha|^2\right),
\label{QPD_1ph}
\end{equation}
which in the limit $s\rightarrow1$ becomes a derivative of
Dirac's $\delta$-function~\cite{VogelBook}:
\begin{eqnarray}
  P_{\rm 1ph}(\alpha)\equiv{\cal W}_{\rm 1ph}^{(1)}(\alpha)=\left(1+
  \frac{\partial}{\partial\alpha}
  \frac{\partial}{\partial\alpha^*}\right) \delta(\alpha),
  \label{Pfock}
  \end{eqnarray}
which can be easily shown by representing $\delta(\alpha)$ as the
limit of the sequence of zero-centered normal distributions, i.e.,
${\cal W}_{\rm vac} ^{(s)}(\alpha)$. Equation~(\ref{Pfock}), and
thus also Eq.~(\ref{QPD_VOPS}) for $s=1$, clearly show a
nonclassical character of any VOPS state (except the vacuum), as
these $P$-functions are more singular than that for a coherent
state $\ket{\alpha}$, i.e., $P_{\rm coh}(\alpha)=\delta(\alpha).$

The Wigner function for an arbitrary VOPS state $\sigma(p,x)$,
which can be obtained from Eq.~(\ref{QPD_VOPS}), reads
\begin{eqnarray}
  W_{\rm vops}(\alpha) &=&
  \frac{2}{\pi}\Big[(1-p) + p(4 |\alpha|^2-1) \nonumber\\
  &&\quad + 2{\rm Re}(x\alpha)
  \Big] \exp\left(-2|\alpha|^2\right).
\label{Wigner}
\end{eqnarray}
Examples of the single-mode Wigner and Cahill-Glauber distribution
for a chosen $\sigma$ state are plotted in Figs.~\ref{fig_Wigner}
and~\ref{fig_Cahill}, respectively.

Experimentally reconstructed two-mode Wigner functions
$W(\alpha_1,\alpha_2)$ are usually shown graphically (see,
e.g.,~\cite{Eichler2011}) via their marginal functions of four
different quadrature pairs, i.e.: $W(X_1,X_2)=\int
W(\alpha_1,\alpha_2) dY_1dY_2$, and analogously $W(Y_1,Y_2)$,
$W(X_1,Y_1)$, and $W(Y_2,X_2)$. As an example, we show such four
marginal distributions in Fig.~\ref{fig_marginals} for
$\rho(p=0.5,x=0.37)$.

The Cahill-Glauber QPD, $W^{(1/2)}(\alpha)$, was calculated using
Eq.~(\ref{multimodeWigner}), while the Wigner functions were
calculated from: (i) the simple formulas in Eqs.~(\ref{QPD_BS2}),
(\ref{QPD_vac}), and (\ref{Wigner}) for the model without phase
damping, i.e., for the BS output state in Eq.~(\ref{RhoOut}); and
(ii) the general definition of the two-mode Wigner function, given
in Eq.~(\ref{QPD_VOPS}), for two-mode states affected by phase
damping according to Eq.~(\ref{RhoOutGen}).

\subsection{Angular-momentum probability surfaces}

Because the studied two-mode states are limited to two qubits or
even formally to a single qutrit, as implied
Eq.~(\ref{RhoOutGen}), we can visualize their properties more
compactly using angular-momentum probability surfaces (AMPS) or,
equivalently, angular-momentum Wigner functions. As defined
in~\cite{Rochester2001, Alexandrov2004, BudkerBook}, an AMPS [say
$\rho_{JJ}(\theta,\phi)$] for a given $(2J+1)\times (2J+1)$ state
$\rho$ (which can be interpreted as an angular-momentum state for
any $J$) is a three-dimensional closed surface, where the distance
from the origin in a specific direction corresponds to the
probability of the maximum projection of $\rho$ along that
direction. An AMPS $\rho_{JJ}(\theta,\phi)$ can be given as a
linear combination of spherical harmonics with the coefficients
corresponding to the moments of a polarization operator, and the
Clebsch-Gordan coefficients (see~\cite{Alexandrov2004} for
details).

A one-to-one correspondence between a given $\rho$ and
$\rho_{JJ}(\theta,\phi)$ can be easily shown by recalling the
orthonormality of the spherical harmonics. Alternatively, one can
apply the angular-momentum Wigner functions introduced by
Agarwal~\cite{Agarwal1981} (see also~\cite{Dowling1994}), which
are also simply related to the AMPS~\cite{BudkerBook}. Thus, the
AMPS, and the above-mentioned standard and generalized Wigner
functions can be interchangeably used as complete representations
of the studied state $\rho$.

In our case, we encode the Fock basis states $\ket{00}$,
$\ket{01}$, and $\ket{10}$ into, respectively,  the angular
momentum states $\ket{J,-1}$, $\ket{J,0}$, and $\ket{J,1}$, where
$J=1$ corresponds to a qutrit. We note that other encodings can
also be applied. We have shown in Fig.~\ref{fig_AMPS}, the AMPS
for chosen states, which reveal different relations between the NC
potentials corresponding to all the hierarchy regimes listed in
Table~I.

\section{Discussion}

\subsection{Experimental feasibility}
\label{Sec_ExperimentalSetups}

\subsubsection{Generation of arbitrary vacuum-one-photon superpositions}

A number of methods for generating superpositions of Fock states,
including the studied VOPS states, have been proposed and
implemented experimentally with optical~\cite{Resch2002,
Lvovsky2002, Lvovsky2002b, Babichev2003, Magro2023} or
microwave~\cite{Hofheinz2009} photons.

In particular, VOPS can be generated from a coherent state by
generalized conditional quantum teleportation and projective
synthesis using a quantum scissors device, as shown schematically
in Fig.~\ref{fig_setup}(b). The method was proposed
in~\cite{Barnett1996}, its experimental feasibility was analyzed
in detail in~\cite{Ozdemir2001, Ozdemir2002}, and it was
experimentally implemented in~\cite{Babichev2003}. The device
comprises two balanced beam splitters BS$_1$ and BS$_2$. A
single-photon state $\ket{1}$ is mixed with the vacuum $\ket{0}$
on BS$_1$, and the generated entangled state at one of the BS$_1$
outputs is mixed with a coherent state $\ket{\alpha}$ (with a
complex amplitude $\alpha$) at BS$_2$. To generate a desired pure
state $\sigma[ p,\sqrt{p(1-p)}]=\ket{\psi}\bra{\psi}$, the
amplitude $\alpha$ should satisfy the condition
$p/(1-p)=|\alpha|^2$, so $\ket{\psi}\sim(\ket{0}+\alpha\ket{1})$.
The projection synthesis of $\ket{\psi} $ is realized by
conditional measurements at the two single-photon detectors, D$_1$
and D$_2$. A proper generation of $\ket{\psi} $ at the second
output port of BS$_1$ occurs if the detector D$_1$ registers a
single photon and D$_2$ does not register any (or vice versa). In
case of other measurement results, the generation (and qubit
teleportation) is unsuccessful, so the procedure should be
repeated. Note that a VOPS state is generated via quantum state
truncation (which can be considered a measurement-induced photon
blockade process) and via the conditional teleportation of the
truncated state.

To generate an incoherent VOPS state $\sigma(p,x)$ with a
coherence factor $|x|<\sqrt{p(1-p)}$, a phase shifter can be
applied (with a specific probability), as shown in
Fig.~\ref{fig_setup}(b). For example, by using random 0 or $\pi$
phase shifts with a given probability, one can decohere a given
pure-state superposition to an arbitrary degree. A phase shifter
can be replaced by two kinds of mirrors changing the phase of a
state during its reflection by either 0 or $\pi$. Let us assume
that the state $\ket{\psi_0}={\cal N} (\ket{0}+\alpha\ket{1})$ for
$\phi=0$ was generated $n_0$ times, and $\ket{\psi_1}={\cal
N}(\ket{0}-\alpha\ket{1})$ for $\phi=\pi$ was produced $n_1$
times, where ${\cal N}$ is the normalization constant. In fact,
the state $\ket{\psi_1}$ is generated in the scheme if a single
photon is detected by $D_2$ instead of $D_1$; thus, no phase
shifter is required for generating $\ket{\psi_1}$. The
corresponding mixed state reads $\sigma'=\sum_{i=0,1}
n_i\ket{\psi_i}\bra{\psi_i}/(n_0+n_1)$; so, if $n_1=n_0$ then
$x=0$, and if $n_1=0$ then $x=\sqrt{p(1-p)}$. Thus, by choosing
properly $n_1$ compared to $n_0$, one can obtain any value of
$|x|\in[0,\sqrt{p(1-p)}]$.

VOPS states can also be generated conditionally (via
postselection) using other linear-optical schemes, e.g., via:
quantum-optical catalysis~\cite{Lvovsky2002}, spontaneous
parametric down-conversion~\cite{Resch2002}, or a single-photon
linear amplification with finite gain~\cite{Ralph2009}.

We focus here on freely propagating VOPS states generated in a
linear-optical system. We note that the generation and control of
arbitrary superpositions of harmonic-oscillator states were
experimentally demonstrated also in various other systems, which
include microwave resonators~\cite{Deleglise2008, Houck2007,
Sillanpaa2007, Hofheinz2009} and optical
cavities~\cite{Boozer2007}, or even ion traps, where
superpositions of motional states of trapped ions were
generated~\cite{Ben-Kish2003}. Thus, our classification of NC is
not limited to VOPS states, but also applies to other bosonic
excitations.

\subsubsection{Two-mode state tomography}

Once a desired VOPS state is generated, it is mixed with the
vacuum on a balanced BS and than a two-mode Wigner function can be
reconstructed using, e.g., homodyne QST as shown in
Fig.~\ref{fig_setup}(a). It should be noted that from the
experimental point of view, it is much more challenging to perform
optical tomography on qubit states implemented as VOPS states
compared to such tomographic measurements of optical qubits
implemented in other ways, including photon polarization. Anyway,
a number of experiments reported the generation of VOPS states and
their tomographic reconstruction via homodyne
detection~\cite{Lvovsky2002, Babichev2003, Babichev2004, Fuwa2015,
Magro2023}. Homodyne tomographic measurements of the joint
detection probabilities for testing Bell nonlocality were first
considered on correlated optical beams at the output of a
nondegenerate parametric amplifier in Ref.~\cite{DAriano1999}.

Thus, a typical setup of two-mode homodyne QST, as schematically
shown in Fig.~\ref{fig_setup}(a), can be applied for
reconstructing a two-qubit Wigner function $W(\alpha_1,\alpha_2)$
from which the corresponding density matrix $\rho_{\exp}$ can be
calculated by Eq.~(\ref{rho_from_QPD}). To find the corresponding
single-qubit state $\sigma(p,x)$, one can numerically find the
closest state $\rho_{qr}(p,x)$, given in Eq.~(\ref{RhoOutGen}),
maximizing the Uhlmann-Jozsa fidelity (or, equivalently,
minimizing the Bures distance),
\begin{eqnarray}
F_{\max}&=&\max_{p,x,q,r} F[\rho_{\rm exp},\rho_{qr}(p,x)]
\nonumber\\
&\equiv& \max_{p,x,q,r} \Big[ {\rm
Tr}\Big(\sqrt{\sqrt{\rho_{\exp}}\rho_{qr}(p,x)\sqrt{\rho_{\exp}}}\Big)
\Big]^2. \label{fidelity}
\end{eqnarray}
Homodyne QST for reconstructing two-mode Wigner function can be
replaced by the Lutterbach-Davidovich QST~\cite{Lutterbach1997}
based on performing proper displacements in a phase space and
parity measurements using the Cahill-Glauber formula, given in
Eq.~(\ref{QPD_tomo}). The single-mode QST method was
experimentally applied in, e.g.,~\cite{Hofheinz2009} for
reconstructing single-mode Wigner functions of Fock-state
superpositions (including VOPS states) in a superconducting
resonator. The Lutterbach-Davidovich method can be readily applied
for reconstructing also two-mode Wigner functions (as
experimentally implemented in, e.g.,~\cite{Wang2016}), in the same
spirit as single-mode homodyne QST was generalized to two-mode QST
[see Fig.~\ref{fig_setup}(a)]. Moreover, a modified
Lutterbach-Davidovich method can be applied for reconstructing
also the single- and two-mode Cahill-Glauber $s$-parametrized QPDs
given in Eqs.~(\ref{QPD}) and~(\ref{Ts2}) for $s$ not too close 1.

The NC of experimental VOPS states can be tested by applying
various NC witnesses, including a Vogel criterion~\cite{VogelBook}
as applied in~\cite{Lvovsky2002}, or negative Wigner
functions~\cite{Magro2023}. The NC of single-photon Fock states
was experimentally tested via violating a Bell inequality
calculated from a two-mode density matrix reconstructed via
homodyne detection in~\cite{Babichev2004, Fuwa2015}; those results
can be considered as a special case of our nonlocality potential
for a single-photon Fock-like state generated experimentally, and
other $\sigma(p,x)$ states were not studied there.

At the end of this section we would like to stress the importance
of applying quantum state tomography in this study. Specifically,
we are interested not only in testing whether a single-mode state
exhibits a given type of quantum correlations, but our goal is to
quantify the NC of the state via measures of two-mode quantum
correlations, and finally to demonstrate the related hierarchy of
such NC quantifiers. This is a much harder problem especially to
determine an entanglement measure of a general two-qubit state
without a full two-qubit QST.  For a related discussion and
references we refer to~\cite{Jirakova2021}, where the hierarchy of
entanglement, steering, and Bell nonlocality of experimental two
polarization qubit states was demonstrated via a full QST.
Actually such a method which enables the determination of an
entanglement measure without full QST of two polarization qubits
has been proposed~\cite{Karol2015b}, but it is quite complicated
and, thus, has not been implemented experimentally yet. The
determination of the Costa-Angolo steering measures $S_{\rm
CA}^{(2)}(\rho)$ and $S_{\rm CA}^{(3)}(\rho)$ (and, thus, the
corresponding Bell nonlocality and steering potentials) without
full QST is possible, but the method has been so far developed
only for polarization qubits~\cite{Karol2017}. To our knowledge,
the only experimental work showing the hierarchy of entanglement,
steering, and Bell nonlocality measures without full QST has been
reported very recently in Ref.~\cite{Abo2023}, but only for some
specific classes of two-polarization qubits (i.e., Werner and
Werner-like states). In the present paper, we study an analogous
hierarchy of quantum correlations, but for single-qubit states.
These states, after mixing with the vacuum on a balanced or
unbalanced BS and subjected to phase damping result in two-qubit
states belonging to much broader classes of states than the Werner
and Werner-like states.

\subsection{Nonclassical potentials for higher-dimensional
single-mode optical states} \label{Sec_CV}

One can apply NC potentials not only for VOPS states, but also for
single-mode optical states of higher dimensions, and (at least for
some classes of) continuous-variable (CV) states. We can interpret
such potentials in close analogy to those for single-qubit states
by applying the Wiseman \emph{et al.} interpretation of the
corresponding two-mode NC correlations~\cite{Wiseman2007}.
Specifically, an EPR steering potential describes the quantum
correlations exhibited by a single-mode bosonic field, enabling
the verification of two-mode entanglement, generated by a linear
coupling of the single-mode field with the vacuum, even when
complete characterization of one of the generated modes is
lacking. While the Bell nonlocality (entanglement) potentials
describe single-mode nonclassical correlations in the case when
complete characterization of both generated modes is lacking
(available).

The calculation of steering and Bell nonlocality potentials based
on measures of the corresponding two-mode correlations would be
very challenging numerically, except low-dimensional qudits or
specific classes of CV states (like Gaussian states). In
particular, the calculation of steering potentials based on
two-mode steering measures for two qutrits can be effectively
performed by applying semidefinite
programming~\cite{CavalcantiReview}. Anyway, such a measure-based
approach becomes numerically demanding already for two quartits.
Thus, it is much more practical to analyze single-mode steering
and nonlocality potentials for qudits and CV systems based on
necessary and sufficient criteria, corresponding to violations of
some classical inequalities for observing two-mode correlations,
instead of analyzing their measures. Thus, the hierarchies of
criteria of  steering and nonlocality potentials for single-mode
fields can be determined via the hierarchies of sufficient or
necessary conditions for observing, respectively, two-mode
steering (e.g.,~\cite{Kogias2015}) and nonlocality
(e.g.,~\cite{Navascus2007}).

The calculations of steering and Bell nonlocality potentials can
usually be much simplified by limiting the number of measurements
from infinite to finite, as we have assumed even in our analysis
of single-qubit states. A variety of powerful Bell and steering
inequalities, which can be readily applied for calculating the
corresponding potentials beyond the VOPS states and beyond the
applied measurement scenarios, are reviewed in
Refs.~\cite{BrunnerReview} and~\cite{CavalcantiReview,UolaReview},
respectively. Steering witnesses for CV systems can be based on
the variances of some observables~\cite{Reid1989} or entropic
uncertainty relations~\cite{Walborn2011,Schneeloch2013}. We also
note that Bell inequalities, which can be the basis for defining
the nonlocality potentials for CV systems, have been studied even
for the infinite number of measurement settings of each
party~\cite{Kaszlikowski1999, Aharon2012} and for continuous sets
of values of the measurement outputs~\cite{Cavalcanti2007,
Salles2010}.

Such an analysis of NC potentials for CV states, can be much
simplified by limiting the interest to Gaussian states, i.e.,
displaced squeezed thermal states. Actually, an entanglement
potential based on the logarithmic negativity was applied to
Gaussian states already in the first paper on NC
potentials~\cite{Asboth2005}. The convertibility (via a BS) of
locally squeezed Gaussian states and entanglement was considered
in Ref.~\cite{Arkhipov2016}. Concerning steering potentials, one
can use a computable measure of steering for arbitrary bipartite
Gaussian states proposed in~\cite{Kogias2015b}. Nonlocality
potentials for Gaussian states can be considered via Bell's
inequality violations using homodyne detection, as studied in,
e.g.,~\cite{Paternostro2009}.

\section{Conclusions} \label{Conclusions}

We have studied theoretically measures of various types of
single-qubit quantum correlations related to two-qubit
correlations via a linear transformation. Thus, we have
generalized the concept of entanglement potentials of Asb\'oth
\textit{et al.}~\cite{Asboth2005}, as measures of single-mode NC,
by proposing the Bell nonlocality and steering potentials.
Analogously to the Wiseman \emph{et al.} standard interpretation
of entanglement, steering, and Bell nonlocality of two-party
systems~\cite{Wiseman2007}, one can interpret NC correlations of
single-qubit states with nonvanishing potentials via trusted or
untrusted detectors used for measuring the two-qubit states, which
are generated via balanced beam-splitting on the single-qubit
ones.


We have applied this approach for quantifying the nonclassicality
of VOPS states by mixing them with the vacuum on a balanced BS and
then to determine various measures of two-qubit (two-mode)
nonclassical correlations. Specifically, we have applied here: (i)
the negativity and concurrence as examples of entanglement
potentials; (ii) quantum steering potentials based on the
Costa-Angelo measures~\cite{Costa2016} of two-qubit steering in
the three-measurement scenario via the maximal violation of a CJWR
inequality. We have chosen these specific steering potentials as
they can be calculated analytically for any two-qubit states. We
note that steering potentials can be defined and applied (at least
numerically) via other popular steering measures, like the
steerable weight~\cite{Skrzypczyk2014} and the steering
robustness~\cite{Piani2015}, which also might be applied for
studying steering potentials for two qudit states. Moreover, we
have defined a Bell nonlocality potential via the Horodecki
measure~\cite{Horodecki1995} of two-qubit Bell nonlocality
quantifying the maximal violation of the Bell-CHSH inequality. We
note that this potential is monotonically related to the steering
potential based on the Costa-Angelo measure in the two-measurement
scenario~\cite{Costa2016}. Thus, with the help of these
potentials, we could reveal the hierarchy of single-qubit
nonclassical correlations in analogy to the hierarchy of the
corresponding two-qubit correlations~\cite{Jirakova2021,Abo2023}.
We have discussed various methods for the generation of VOPS
states and the homodyne tomographic reconstruction of the
resulting two-mode states and the calculation of realistic
potentials assuming system imperfections including phase damping
and unbalanced beam splitting.

The studied hierarchy of single-qubit potentials for generating
two-qubit entanglement, steering, and Bell nonlocality can also be
useful for estimating the degree of one type of quantum
correlation from another, e.g., estimating the Bell nonlocality or
steering potentials from an entanglement potential (or vice
versa), in the spirit of such estimations for the corresponding
two-qubit quantum-correlation measures (see,
e.g.,~\cite{Karol2013} and references therein).

Apart from a fundamental interest in single-photon entanglement
and VOPS states, these have been experimentally used for quantum
information tasks, including quantum
teleportation~\cite{Lombardi2002,Babichev2003} and EPR
steering~\cite{Fuwa2015}. Moreover, one can subject a VOPS to a
non-demolition photon presence detection gate and to partially
erase this information~\cite{Roik2020}. Thus, we believe that a
deeper study NC correlations of VOPS states can find further
applications for quantum technologies.

We also stress that the studied NC potentials are not limited to
Fock-state superpositions. Indeed, the results of this paper can
be experimentally implemented with qubits encoded in, e.g., photon
polarization, as reported in Ref.~\cite{Kadlec2023}. Thus, we
believe that our work can stimulate further research in
quantifying and utilizing the NC of single-mode optical fields in
close analogy to various types of intermode quantum correlations
with applications for quantum information processing.

\acknowledgements

A.M. and K.B. are supported by the Polish National Science Centre
(NCN) under the Maestro Grant No. DEC-2019/34/A/ST2/00081. J.K.
acknowledges Internal Palack\'y University grant No.
IGA\_PrF\_2023\_005. F.N. is supported in part by: Nippon
Telegraph and Telephone Corporation (NTT) Research, the Japan
Science and Technology Agency (JST) [via the Quantum Leap Flagship
Program (Q-LEAP), and the Moonshot R\&D Grant Number JPMJMS2061],
the Asian Office of Aerospace Research and Development (AOARD)
(via Grant No. FA2386-20-1-4069), and the Foundational Questions
Institute Fund (FQXi) via Grant No. FQXi-IAF19-06.

\appendix

\section{Cahill-Glauber $s$-parametrized quasiprobability distributions}
\label{Appendix_Wigner}

Here we recall some basic formulas of the Cahill-Glauber formalism
of quasiprobability distributions (QPD)~\cite{Cahill1969}, which
are phase-space representations of single- or multimode states.
These are generalizations of the standard Wigner, Husimi, and
Glauber-Sudarshan functions.

For a multimode optical state $\rho$, the Cahill-Glauber
$s$-parametrized QPD is defined as
\begin{equation}
{\cal W}^{(s)}(\{\alpha_k\}) =  \frac1{\pi^M}\langle
{T}^{(s)}(\{\alpha_k\})\rangle= \frac1{\pi^M}{\rm Tr} \Big[ {\rho}\: \prod_{k}
{T}^{(s)}(\alpha_k) \Big], \label{QPD}
\end{equation}
where $s\in[-1,1]$, $\{\alpha_k\}=(\alpha_1,...,\alpha_M)$ (for
the studied states $\rho_{\rm out}$, the number of modes is
$M=2$),  $\alpha_k$  are complex numbers, and the $k$th-mode
operator $T^{(s)}(\alpha_k)$ is defined by
\begin{equation}
  {T}^{(s)}(\alpha_k) =
\int{D}^{(s)}(\beta_k)\: \exp\!\left( \alpha_k\beta_k^*-
\alpha_k^*\beta_k\right) \frac{{\rm d}^2 \beta_k}{\pi},
\label{Ts1}
\end{equation}
which is the Fourier transform of the $s$-parametrized
displacement operator,
\begin{eqnarray}
{D}^{(s)} (\beta_k)=\exp \left( \beta_k {a}_k^\dagger - \beta_k^*
{a}_k +\frac{s}{2} \left| \beta_k \right|^2 \right), \label{Ds}
\end{eqnarray}
where $a_k$ ($a_k^\dagger$) is the $k$th-mode annihilation
(creation) operator. The multimode operator
${T}^{(s)}(\{\alpha_k\})$ is just a product of single-mode
operators ${T}^{(s)}(\alpha_k)$, which can be equivalently defined
as
\begin{eqnarray}
  {T}^{(s)}(\alpha_k) = \frac{2}{1-s} D(\alpha_k)
  \left(\frac{s+1}{1-s}\right)^{a_k^\dagger a_k} D^{-1}(\alpha_k),
\label{Ts2}
\end{eqnarray}
where $D(\alpha_k)={D}^{(0)} (\alpha_k)$ is the standard
displacement operator. In the three special cases of $s=-1,0,1$,
the $s$-parameterized QPD, ${\cal W}^{(s)}(\alpha_1,\alpha_2),$
reduces, respectively, to the Husimi $Q$, Wigner $W$, and
Glauber-Sudarshan $P$ functions corresponding to the antinormal,
symmetric, and normal orderings of the creation and annihilation
operators. After substituting Eq.~(\ref{Ts2}) to Eq.~(\ref{QPD})
for $s=0$, one arrives at
\begin{eqnarray}
  W(\{\alpha_k\})&\equiv&W^{(0)}(\{\alpha_k\})\\
  &=&\frac2{\pi^M} \tr\Big[\rho\prod_k D(\alpha_k)\mathcal{P}(a_k)
  D^{-1}(\alpha_k)\Big],\nonumber
  \label{QPD_tomo}
\end{eqnarray}
where $\mathcal{P}(a_k)=(-1)^{a_k^\dagger a_k}$ is the
photon-number parity operator. The Cahill-Glauber formula in
Eq.~(\ref{QPD_tomo}) is the basis for a direct experimental
measurement of the single-mode~\cite{Lutterbach1997, Hofheinz2009}
and multimode Wigner functions just by performing proper
displacements $D(\alpha_k)$ in the phase space and the
measurements of the parity operator $\mathcal{P}(a_k)$.

The QPD for any $s$ contains a full information about a given
state ${\rho}$, as implied by the formula
\begin{eqnarray}
{\rho} &=& \int{\cal W}^{(s)}(\{\alpha_k\})\:
{T}^{(-s)}(\{\alpha_k\})\: {\rm d}^2\{\alpha_k\},
\label{rho_from_QPD}
\end{eqnarray}
where ${\rm d}^2\{\alpha_k/\pi\}={\rm d}^2\alpha_1\cdots {\rm
d}^2\alpha_M$. For numerical calculations of a QPD (practically
for  any $s$, which is not too close to 1), it is useful to use
its Fock-state representation,
\begin{eqnarray} {\cal
W}^{(s)}(\{\alpha_k\}) =\frac1{\pi^M}
\sum_{\{m_k\}=0}^{N_{0}}\:\:\sum_{\{n_k\}=0}^{N_{0}} \prod_{k=1}^M \langle n_k| {T}^{(s)}(\alpha_k) |m_k \rangle, \nonumber\\
\times \langle  \{n_k\}|{\rho}|\{m_k\},\rangle\hspace{3cm}
\label{multimodeWigner}
\end{eqnarray}
where
\begin{eqnarray}
\langle n_k| {T}^{(s)}(\alpha_k) |m_k \rangle =
\sqrt{\frac{n_k!}{m_k!}} \left(
\frac{-s_m}{s_p}\right)^{n_k}s_m^{\delta_k+1}
\left(\alpha_{k}^{*}\right)^{\delta_k} \nonumber\\ \times
L_{n_k}^{(\delta_k)} \left(s_p s_m|\alpha_k|^2 \right)
\exp\left(-s_m|\alpha_k|^2 \right) ,\quad\quad \label{Tnm}
\end{eqnarray}
for $m_k\ge n_k$; other elements can be found from the property
$\langle n_k| {T}^{(s)}(\alpha_k) |m_k \rangle=\langle m_k|
{T}^{(s)}(\alpha^*_k) |n_k \rangle$. Here $\delta_k=m_k-n_k$,
$s_{p}=2/(1+ s)$, $s_{m}=2/(1- s)$, and $L_{n_k}^{(\delta_k)}(x)$
are the associated Laguerre polynomials. To calculate the QPD for
a given $s<1$, we can directly apply the density matrices given in
Eqs.~(\ref{RhoOut}) and~(\ref{RhoOutGen}) to Eq.~(\ref{Tnm}). The
formula in Eq.~(\ref{Tnm}) can be applied even in the limit
$s\rightarrow 1$, but the limit should be taken very carefully.

We note that for the VOPS states $\sigma$ and the BS-transformed
states $\rho_{\rm out}$, it is enough to analyze the two special
cases of the polynomials: $L_{0}^{(\delta_k)}(x)=1$ and
$L_{1}^{(\delta_k)}(x)=1+\delta_k-x$, because $N_{0}=1$. Thus, by
denoting $T^{(s)}_{nm}(\alpha)=\langle n| {T}^{(s)}(\alpha) |m
\rangle$, we have
\begin{eqnarray}
  T^{(s)}_{00}(\alpha) &=& \frac{2}{1- s} \exp\left(-\frac{2}{1- s} |\alpha|^2\right), \nonumber \\
  T^{(s)}_{10}(\alpha) &=& [T^{(s)}_{01}(\alpha)]^* = \frac{4\alpha}{(1- s)^{2}}  \exp\left(-\frac{2}{1- s} |\alpha|^2\right), \nonumber \\
  T^{(s)}_{11}(\alpha) &=& \frac{2(4 |\alpha|^2+s^2-1) }{(1- s)^{3}}\exp\left(-\frac{2}{1- s} |\alpha|^2\right).\quad
\label{Tnm_VOPS}
\end{eqnarray}


\end{document}